\documentclass[%
 reprint,
showpacs,
preprintnumbers,
 amsmath,amssymb,
 aps,
]{revtex4-1}

\usepackage{graphicx}% Include figure files
\usepackage{dcolumn}% Align table columns on decimal point
\usepackage{bm}% bold math
\usepackage[T1]{fontenc}
\usepackage{textcomp}

\usepackage{amsfonts}
\usepackage{upgreek}     % for non italic greek symbols
\usepackage[T1]{fontenc} % for accents etc
\usepackage{ae,aecompl}  % for pdf fonts

\newcommand{\micron}{~\ensuremath{\upmu\text{m}}}
\newcommand{\umns}{\ensuremath{\upmu\text{m}}}

\begin{document}

\title{Trapping molecules on chips}
\author{G. Santambrogio}
\email{santambrogio@lens.unifi.it}
\affiliation{Fritz-Haber-Institut der Max-Planck-Gesellschaft,
Faradayweg 4-6, 14195 Berlin, Germany.\\
Istituto Nazionale di Ottica-CNR \& European Laboratory for Non-Linear Spectroscopy
LENS, Via Nello Carrara 1, 50019 Sesto Fiorentino, Italy.}
\date{\today}

\begin{abstract}
In the last years, it was demonstrated that neutral molecules can be
loaded on a microchip directly from a supersonic beam. The molecules
are confined in microscopic traps that can be moved smoothly over
the surface of the chip. Once the
molecules are trapped, they can be decelerated to a standstill, for
instance, or pumped into selected quantum states by laser light or
microwaves. Molecules are detected on the chip by
time-resolved spatial imaging, which allows for the study of the
distribution in the phase space of the molecular ensemble. 

\end{abstract}

\maketitle

\section{Introduction}

In several fields, miniaturization improved the performances of the
devices and proved itself economically convenient. For chemistry,
miniaturized devices shrink the pipettes, beakers, and test tubes of a
modern lab onto a microchip-sized
substrate.\cite{Daw_Nature442p367_2006}  Lab-On-a-Chip (LOC)
technology exploits the progresses in microfluidics, enjoying very
fast transport times and accurate knowledge of molecular
concentrations. This in turns makes analysis both faster and more
accurate. Moreover, instead of carrying the samples to be analyzed to
a central laboratory, LOCs are deployed directly on the field, with
applications from the international space
station~\cite{Morris_Astrobiology12p830_2012} to
antiterrorism~\cite{Frisk_LabChip6p1504_2006}. Economical convenience
stems both from the low production costs of LOC devices and from
savings in reagents costs and waist disposal.  

At the present stage of development, however, there is a natural limit
to the level of control on the reaction parameters. If the knowledge
of molecular concentration is to be extended to the level of single
molecules and the interaction energy enhanced to the mK level, one
cannot ignore perturbations due to physical and chemical effects of
the chip itself, like capillary forces and chemical interactions of
the construction materials. One possible solution involves avoiding
any direct contact of the chemical species under investigation with the
substrate of the chip itself. Infrared spectroscopy in the gas phase
is currently used on cold (a few K) ionic species to study solvation,
by adding the molecules of solvent one by
one~\cite{Heine_IntRevPhysChem34p1_2015,Asmis_AccChemRes45p43_2012,Johnson_JChemPhys139p224305_2013,Fournier_Science344p1009_2014}. Moreover,
reactivity studies on clusters that mimic the interplay between
substrate and active species in heterogeneous 
catalysis~\cite{Jiang_JPhysChemA115p11187_2011,Janssens_PhysRevLett96p233401_2006}
have become possible. One might thus conceive a future in which a 
countable number of molecules, possibly with their solvation shell(s),
will be manipulated with electromagnetic fields above the surface of
the chip and the chemical analysis will reach ultimate accuracy.

Another field that greatly benefited from miniaturization is atomic
physics. Two ingredients lead to the success of atom chips. One is the
efficient laser cooling of atoms~\cite{Metcalf_LaserCooling1999}. The
other is the notion that miniaturization of magnetic field structures
enables the creation of large field gradients, i.e., large forces and
steep potential wells. Today, the manipulation of atoms above a chip
using magnetic fields produced by current-carrying wires is a mature
field of research.\cite{Fortagh_RevModPhys79p235_2007} Such atom chips
have been used to demonstrate rapid Bose-Einstein
condensation~\cite{Hansel_Nature413p498_2001} and have found
applications in matter-wave interferometry and in inertial and
gravitational field
sensing~\cite{Schumm_NatPhysics1p57_2005,Zoest_Science328p1540_2010},
quantum computation~\cite{Ospelkaus_Nature476p181_2011}, and many-body
nonequilibrium physics~\cite{Gring_Science337p1318_2012}. 

Molecules are not only the building blocks of chemistry and the
natural extension of atomic physics, i.e. a bridge between fundamental
quantum physics and the richness of the chemical world. With their
numerous internal degrees of freedom and strong long-range
interactions, they are ideal systems for the investigation of
fundamental phenomena.  Molecules are used for the measurement of the
electron electric dipole moment~\cite{Baron_Science343p269_2014},
measurements of parity violation in chiral molecules~\cite{Daussy_PhysRevLett83p1554_1999}, tests of fifth
forces~\cite{Salumbides_PhysRevD87p112008_2013} and
QED~\cite{Salumbides_PhysRevLett107p043005_2011}, and
measurements of fundamental constants~\cite{Daussy_PhysRevLett98p250801_2007} and their possible
variation~\cite{Shelkovnikov_PRL100p150801_2008,Truppe_NatureComm4p2600_2013,Bagdonaite_Science339p46_2013,Salumbides_PhysRevLett101p223001_2008}.
Molecules allow unique approaches to quantum
computation~\cite{DeMille_PhysRevLett88p067901_2002,Andre_NatPhys2p636_2006} and 
can condense to new quantum
phases~\cite{Goral_PhysRevLett88p170406_2002,Micheli_NaturePhys2p341_2006}.
Moreover, novel quantum-mechanical collision and reaction channels are
predicted for cold molecules~\cite{Krems_PhysChemChemPhys10p4079_2008}, where
field-induced alignment~\cite{deMiranda_NaturePhys7p502_2011} and
field-sensitive collision
resonances~\cite{Avdeenkov_PhysRevA66p052718_2002} allow for the study of
controlled chemistry~\cite{Tscherbul_JChemPhys129p034112_2008}.

Here, the techniques to trap cold molecules on microchips are introduced
and the recent developments in this field are reviewed. First, the essential
features of microchip design and the necessary experimental setup are
described. Then, the problem of non-adiabatic losses from the
microtraps is addressed and the most viable solutions are
presented. Further, some recent results on state transition of trapped
molecules are presented, involving rotational and vibrational
transitions. And finally, on-chip detection and imaging is briefly
discussed. 

\section{Microchip design and experimental setup}
In contrast to ultracold atoms, for which efficient cooling was
realized early on~\cite{Metcalf_LaserCooling1999}, the complicated
level structures of molecules result in a general lack of closed
two-level systems that are necessary for efficient laser
cooling. Therefore, laser cooling~\cite{Shuman_Nature467p820_2010} and
slowing~\cite{Barry_PhysRevLett108p103002_2012} of molecules is
currently limited to a few
species~\cite{Zhelyazkova_PhysRevA89p053416_2014} and the temperature
so-far achieved are in the order of a few mK. Instead, the most
versatile and intense sources of cold molecules are cryogenic
buffer-gas cooling~\cite{Weinstein_Nature395p148_1998} and supersonic
expansions. These methods deliver a relatively cold sample molecules,
$\sim1$~K, albeit with a large velocity in the laboratory frame,
100--400~m/s.

\begin{figure}
\centering
\includegraphics[width=0.45\textwidth]{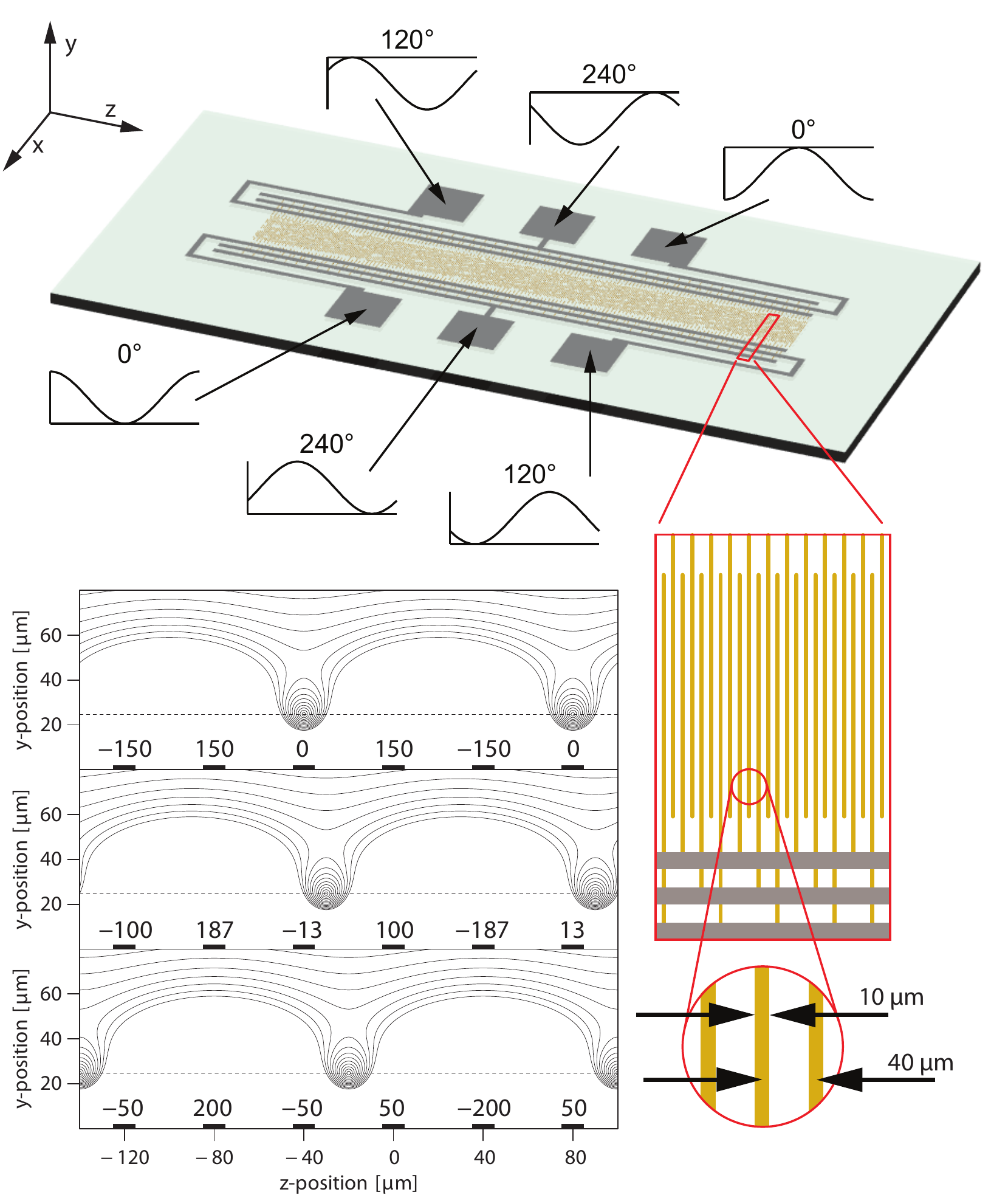}
\caption{Picture of the microchip used by Meek et
  al.\cite{Meek_NewJPhys11p055024_2009,Meek_PhysRevLett100p153003_2008}. The
  structure of the electrodes is shown in three stages of
  magnification. The $y$ axis is here chosen perpendicular to the
  substrate, the $z$ axis is along the substrate, perpendicular to the
  electrodes. Calculated contour lines of equal electric field
  strength (intervals of 0.5~kV/cm) are shown in the bottom-left part
  of the Figure for the ($z$-$y$) plane, above the periodic array of
  electrodes. The sections of the electrodes are represented as black
  rectangular boxes. The values of the applied potentials in V are
  given for three different times in the harmonic waveform cycle.  
\label{fig01}} 
\end{figure}

Not everything is harder with molecules than with atoms, though. In
fact, polar molecules, are sensitive enough to electric fields that
electric trapping is relatively easy~\cite{Bethlem_Nature406p491_2000}
and it has been demonstrated even for molecules without any
cooling~\cite{Rieger_PRL95p173002_2005}. 
Electric manipulation of polar molecules is based on the Stark shift
of quantum levels in the presence of an electric
field.~\cite{Bethlem_IntRevPhysChem22p73_2003} The magnitude
of the Stark shift of a certain level is given by $-\mu\,E$, where $E$
is the magnitude of the electric field, and $\mu$ is the mean value of
the component of the electric dipole moment along the direction of the
field. If $\mu$ is antiparallel to the field for a given quantum
state, the level's energy increases with increasing electric field
strength and the state is then called a low-field seeker (lfs). Vice-versa,
if the energy of the quantum state decreases with increasing strength
of the electric field, the state is called a high-field
seeker (hfs).\cite{Meerakker_ChemRev112p4828_2012} Around
the degeneracy at $E=0$, the direction of the electric dipole moment is
not defined and non-adiabatic transitions between states are thus possible.
Since an electric field strength maximum cannot be realized in the
free space, trapping of polar molecules on chips has been demonstrated
only for lfs. It is worth noting that this is a
significant limitation because the ground state of every molecule is
always a hfs.

A design for trapping molecules on a microchip was first presented in
2008 by Meek et al.~\cite{Meek_PhysRevLett100p153003_2008} and only minor improvements have been done ever since. The
operation principle relies on the superposition of electric fields
created by the electrodes on the chip. When two dipolar fields with
different length scales and opposite directions are superimposed, a
minimum of the electric field strength is created. The minimum is
located at the point where the long-range dipole that dominates far
from the surface is canceled by the short-range dipole that dominates
close to the surface. Such a minimum in the electric field strength is
a trap for lfs.\cite{Meek_NewJPhys11p055024_2009} A picture of the
chip used by Meek et al.\ is shown in Fig.~\ref{fig01}. The
electrode design consists in an array of equidistant, parallel, gold
electrodes, each of which is 4~mm long, 10\micron\ wide and
approximately 300~nm high. They are deposited onto a 1~mm thick glass
substrate with a center-to-center spacing of 40\micron\ and are coated
with a 5-\micron\ layer of insulating SU-8. This structure
is periodically extended over about 50~mm. Each electrode is
electrically connected to the electrodes that are (multiple of) six
positions further, i.e.\ the electric field repeats itself every
240\micron. This geometry requires the metallization on the chip to be
deposited on three different levels with insulating SU-8 layers
in-between. The 4-mm long electrodes are all on the same
plane, whereas the connections to the distribution buses are on
different levels. When appropriate potentials are applied to the electrodes, 
an array of tubular electric field geometries of 4~mm length and
20\micron\ diameter are generated, on the axis of which the electric
field strength drops to zero. These electric field geometries act
as traps for lfs and are centered roughly 25\micron\ above the chip surface.
In the bottom-left portion of the figure, the calculated
contour lines of equal electric field strength above the chip are
shown. The horizontal black boxes represent the section of the
electrodes. The potentials applied to the electrodes are indicated
directly above them. In the bottom panel, the situation is shown in
which the electrodes at $z=-40$\micron\ and $z=0$\micron\ create an electric field
that is parallel to the $z$ axis, in the direction of negative $z$
values, for any point along the vertical line at $z=-20$\micron, whereas the
electrodes at $z=40$\micron\ and $z=-80$\micron\ create an electric field that is
pointing in the opposite direction for any point along this
line. Close to the substrate, the field due to the two nearest
electrodes dominates, whereas further away the field
due to the next nearest electrodes is most important. It is clear,
therefore, that at some point above the substrate, at a typical height
on the order of the distance between adjacent electrodes, a zero of
the electric field strength will be generated on the
$z=-20$\micron\ axis. At the ends of the central array in the $\pm
x$-directions, there is a jump
from a region where the potentials are on average zero to a region
where the potentials are on average positive or negative (see
Fig.~\ref{fig01}). This produces an electric field along the
$x$-direction that closes the tubular traps at the two sides. However,
the molecules interact only very seldom with the traps ends because of
the extreme aspect ratio of these tubular traps. 

Using different sets of potentials, it is also possible to position
the minima either directly above an electrode (top panel), or in an
intermediate position (central panel). In fact, by applying the
appropriate potentials, the minima can be positioned at any
$z$-position within the 240-\umns\ period, while their $y$-position
remains constant. Further, the periodic arrangement of the electrodes allows
for a continuous movement of the electric field minima outside the
240-\umns\ period, that is over the macroscopic distance of
the whole device at a constant height of about 
25\micron. For a given trap strength, there is a bijective
relationship between the set of the applied potentials and the
position of the microtraps array, so that the applied potentials must
be periodic functions when the field minima travel across multiple
periods. For the electrodes configuration chosen by Meek et al., 
the applied potentials that generate the smoothly moving traps are six
approximately harmonic functions. Three of the potentials can always be positive,
the other three always negative, and within each polarity set the
potentials need to be phase shifted by $2\pi/3$. Since two microtraps
are formed per period, as can be seen in Fig.~\ref{fig01}, the
effective periodicity is of 120\micron. Thus, the microtraps move at
300~m/s when the sinusoidal modulation of the potential has a
frequency of 2.5~MHz. The trap depth obtained with sinusoidal modulations of 200-V
amplitude peak-to-peak is of the order of 5~kV/cm. For CO molecules in
the upper $\Lambda$-doublet component of the $a^3\Pi_1$, $v=0$, $J=1$
state, for instance, this electric field strength corresponds to a thermal
energy of about 70~mK.

Polar molecules flying parallel to the surface of the chip along the
$z$-direction can be trapped in the microtraps directly from a
supersonic molecular beam.~\cite{Meek_PhysRevLett100p153003_2008} The
density of the molecular beam at the 
chip entrance determines the density of trapped molecules, which is
typically in the order of 10$^7$ molecules/cm$^3$.
Initially, the frequency of the applied
waveforms is chosen to match the velocity of the microtraps to that of
the incoming molecules. Then, the trapped molecules can be brought to
a standstill, or to any intermediate velocity, by continuously
reducing the frequency of the waveforms. An acceleration of the order of
10$^5$~g can be applied, allowing to stop a supersonic beam within a
few cm. 

%Figure~\ref{fig02}(a) shows the signal measured after the chip by
%photo-ionizing the molecules and collecting the ions. To measure the
%free molecular beam, the waveforms are turned off; when the waveforms
%are turned on, a portion of the molecular beam is trapped and, in this
%case, guided at constant speed along the chip

Moving away from the substrate, the contour lines run ever more
parallel to the surface and the strength of the electric field decays
exponentially with the $y$-position. In the region far away from the
surface, therefore, the electrode array yields a flat, repulsive
mechanical potential for polar molecules in lfs states: a
mirror. Indeed, an electrostatic mirror consisting of an array of
parallel and equidistant electrodes on a surface to which alternating
voltages are applied was first discussed by Opat et
al.~\cite{Opat_ApplPhysB54p396_1992}. Based on this principle, both
plane~\cite{Schulze_PhysRevLett93p020406_2004} and
focusing~\cite{GonzalezFlorez_PhysChemChemPhys13p18830_2011}
microstructured mirrors for polar molecules have been
realized. Moreover, Englert et
al.~\cite{Englert_PhysRevLett107p263003_2011} placed two such
microstructured mirrors facing each other as the two faces of a
capacitor to create a macroscopic (2 by 3~cm) electrostatic trap. To
achieve transverse confinement of the molecules they used a
high-voltage electrode between the plates that surrounds the perimeter
of the trap. 

Imaging experiments of the molecular beam with the focusing microstructured
mirror~\cite{GonzalezFlorez_PhysChemChemPhys13p18830_2011} and of
trapped molecules above the
chip~\cite{Marx_PhysRevLett111p243007_2013} allow to estimate the 
effects of the charge that accumulates on the SU-8 insulating layer
above the electrodes. When all electrical potential are not symmetric
about ground, it was found that the mechanical potential becomes
weaker toward the distribution buses that are on average further away
from ground. This is interpreted as the effect of charge that
accumulates on the dielectric, screening the electric field of the
electrodes. Therefore, although the molecules on the chip should in
principle only be sensitive to the electric field strength, the
values of the applied potentials with respect to ground turn out to be
also important. Moreover, the amount of trapped molecules increases
with increasing amplitude of the applied potentials until about
240~V. This limit is probably an effect of the suface charges and is
usually reached without damages to the microstructure.

\section{Non-adiabatic losses}

Thus far we have been assuming that the force imposed on the molecules
only depends on the gradient of the electric-field strength but not on
the direction of the field itself. This is usually a good assumption,
since the molecules reorient themselves and follow the new
quantization axis when the field changes direction, and their potential
energy changes smoothly with the strength of the field. This
approximation can break down, however, when the quantum state that is
used for manipulation couples to another quantum state that is close
in energy. If the energy of the quantum state or the field direction
changes at a rate that 
is fast compared to the energetic splitting, transitions between these
states are likely to occur. Such transitions are particularly
disastrous if lfs end up as hfs or in states that are only weakly
influenced by the electric fields, as this results in a loss of the
molecules from the trap. For atoms in magnetic traps, such losses are
known as Majorana spin-flip transitions. Both for atoms in magnetic
quadrupole traps~\cite{Petrich_PhysRevLett74p3352_1995} and for polar
molecules in electric quadrupole
traps~\cite{Kirste_PhysRevA79p051401R_2009} it has been shown that
non-adiabatic losses are inversely proportional to the square of the
diameter of the particle cloud. A straightforward solution to avoid
Majorana transitions involves the use of an offset
magnetic~\cite{Fortagh_RevModPhys79p235_2007} or
electric~\cite{Kirste_PhysRevA79p051401R_2009} field. Due to the
geometry of the molecule chip, however, applying a static offset
electric field is much harder. In particular, it cannot be done
without leaving the two dimensions of the present
devices~\cite{Meek_PhysRevA71p065402_2005} and the field generated by
external electrodes perpendicular to the substrate would be strongly
screened by the metallic surfaces of the 
microelectrodes, which are only a few microns away from the
molecules. Thus other solutions must be sought.

It was demonstrated in the cases of ammonia~\cite{Kirste_PhysRevA79p051401R_2009} 
and of carbon monoxide~\cite{Meek_Science324p1699_2009}, for instance,
that the choice of the appropriate isotopologue can induce a
beneficial separation of the levels at zero electric field, thus
reducing the non-adiabatic losses. For the simpler case of CO in the
$a^3\Pi$ state, a degeneracy at zero field between two
low-field-seeking levels and a level that is not sensitive to electric
field in $^{12}$CO is lifted in $^{13}$CO due to hyperfine splitting
(the $^{13}$C nucleus has a nuclear spin $|\vec{I}| = 1/2$), and the
low-field-seeking levels never come closer than 50~MHz to the
nontrappable levels. An alternative solution was demonstrated for CO
and consists in the use of a magnetic
field.~\cite{Meek_PhysRevA83p033413_2011} If a magnetic field is
applied in addition to the electric field, a splitting can be induced
between the low-field-seeking and the nontrappable levels of $^{12}$CO
that depends on the strength of the applied magnetic field.  

The solutions mentioned in the last paragraph act on the level
splitting. But the rate at which the energy of the levels changes when
the molecules travel across the traps must also be considered, because
it is the ratio between these two that 
determines the transition probability. Since the trapping potential on
the molecule chip is obtained as a difference between large electric
fields, it turns out to be very sensitive to imperfections in the
applied waveforms. As a result, the microtraps can be jittering around
much faster than the velocity of the trapped molecules with respect to
the averaged trap center. Thus, improving the quality of the waveforms
is a further way to reduce non-adiabatic transitions. Although this is
a non-trivial task because amplitudes over 200~V between 3~MHz and DC
are needed, the improvements done thus far are encouraging. A first
reduction of the waveforms anharmonicity from 7\% to 3\% reduced the
non-adiabatic losses by 10\% and halved the magnitude of the magnetic
field required to saturate the loss
suppression.~\cite{Meek_PhysRevA83p033413_2011} 

The suppression of non-adiabatic losses is a strong motivation to
improve the quality of the waveforms. However, wide bandwith and low
anharmonicity of the amplifiers that generate the waveforms turn out
to be crucial for shutting down the electric field rapidly and
accurately for the imaging experiments (see below). The first
generation of the amplifiers is based on a AB design, realized with
vacuum tubes. The second generation is a class-A design, realized with
semiconductor transistors. The frequency response of each amplifier
is measured and fed back to the software that calculate the
waveform. The current state of the art is a total harmonic distortion
below $-43$~dB.

\section{Addressing state transitions in trapped molecules}
Molecules on a chip can be coupled to photons over a wider range of
frequencies than atoms. The fundamental molecular vibrational modes
can be addressed with mid-infrared photons whereas their overtones and
combination modes extend into the near-infrared range. In addition,
polar molecules have a dense set of rotational transitions in the
sub-THz, or mm-wave, region of the spectrum. Being able to induce a
transition to another internal quantum state in the molecule is
particularly interesting when the molecule remains trapped in the
final state as well.  

Abel et al.~\cite{Abel_MolecularPhysics110p1829_2012} showed an
effective vibrational population transfer in molecules trapped on a
chip. They coupled pulsed IR radiation around 5.9\micron\ to transfer
trapped CO molecules in the $a^3\Pi_1$ state from the $v=0$ to the
$v=1$ levels and addressed both Q- and R-branch
transitions. Crucially, the Stark broadening of the vibrational
transitions induced by the trapping fields is comparable with the
laser bandwidth of about 2.5~GHz and therefore virtually all molecules
could be addressed by the IR radiation. The situation is different
with rotational transitions. First, rotational transitions are
usually more sensitive to electric fields and the inhomogeneous
broadening in the traps is larger than for vibrations. Second,
sub-THz sources are typically narrow. Therefore, when rotational
transitions were addressed in molecules on the
chip~\cite{Santambrogio_ChemPhysChem12p1799_2011}, the microtraps had
to be switched off temporarily to allow for effective rotational
pumping. Interestingly, molecules could be recaptured after being
transferred between different rotational levels. These results thus
demonstrate that mm-wavelength radiation can be coupled to CO
molecules located at less than 50\micron\ above the surface of the
chip---a few hundredths of the wavelength---and rotational spectra
were obtained with a resolution of approximately half a MHz. An
external magnetic field was used to split the Zeeman components and
the resolution of the spectrum is limited by the 
inhomogeneities of the magnetic field above the chip induced by the
metal components of the chip holder.

\section{Imaging molecules on the chip}

The lack of closed two-level systems that makes laser cooling so hard
for molecules is also responsible for the difficulties in molecular
detection using absorption or laser-induced fluorescence. Moreover, in
the presence of a physical structure such as a microchip, scattering
or laser-induced fluorescence from surfaces adds noise to images,
something which is critical when working with small samples. For these
reasons, on-chip detection has only recently been
demonstrated.~\cite{Marx_PhysRevLett111p243007_2013} On-chip detection 
is based on resonance-enhanced multiphoton ionization
(REMPI)~\cite{Ashfold_AnnuRevPhysChem45p57_1994}, which is quantum
state selective, is intrinsically background-free, and is of general
applicability. REMPI is obtained on the chip by illuminating the
molecules with a sheet of light parallel to the surface of the
chip. It is worth noting that ions are several orders of magnitude
more sensitive to electric fields than polar molecules. However, if
care is taken to carefully zero all electric fields used to manipulate
the neutral molecules before the ions are created, it is possible to
create a spatial image of the molecules in a microchannel-plates
detector. A set of magnifying ion lenses can be used to resolve the
molecular distribution in individual microtraps, and since the timing
of the ionizing laser can be tuned, one can follow the spatial
evolution of the molecular clouds in time.  

With this setup for time-resolved spatial imaging, it was possible to
study the phase-space distribution of trapped molecules. The
experiments are done in a similar fashion as for cold atoms: the traps
are quickly turned off and a series of snapshots at different times
return a movie of the ballistic expansion of the particles. In
particular, this allows for a direct measurement of the temperature of
the trapped molecules. Moreover, the cooling induced on the molecular
ensemble by an adiabatic expansion process, induced by slowly
weakening the trapping potential, can be clearly seen to lower the
temperature of the molecules to about a third of the initial value
(from 16 to 5~mK).\cite{Marx_PhysRevLett111p243007_2013}

\section{Outlook}

Miniaturization yields large forces (e.g.~10$^5$~g$\cdot$28 amu, for
CO) at the moderate electric field strenghts of a few kV/cm, with applied
potentials of the order of $\pm$200~V. For comparisons, macroscopic Stark
decelerators achieve accelerations about ten times lower, with electric
field strenghts in the order of 100~kV/cm, and applied voltages in the
10~kV range. Microchip-based devices
might thus be promising for the manipulation of heavy molecules that are
typically only low-field seeking when the electric field is small.~\cite{Meerakker_ChemRev112p4828_2012,Bulleid_PhysRevA86p021404R_2012}

To confine molecular samples with a temperature of about 10~mK, the
Stark broadening induced by the inhomogeneous trapping fields is in
the GHz range. This precludes the application of traps
in precision spectroscopy and collision experiments. For
molecule-based spectroscopic measurements, the shot-noise limit on the
statistical error is proportional to $1/\tau \sqrt N$, where $\tau$ is
the time each molecule spends in the light field and $N$ is the total
number of molecules that participate in the experiment. Moreover, the
major sources of systematic errors and uncertainties in molecular
spectroscopy are the presence of stray fields and Doppler
broadening. These two issues are usually approached by keeping the 
size of the experimental apparatus small, reducing the magnitude of
all necessary fields, and by employing Doppler-free techniques, like
for instance two photons spectroscopy. Therefore, the use of a
microchip to produce a slow beam of cold molecules for spectroscopy is not necessarily as 
inconvenient as it might seem at a first glance. First, the density of trapped
and decelerated molecules on the microchip is the same as for
macroscopic devices or for free molecular beams---only the absolute
number of molecules is small. Since a minimum of electric field is
required for a population transfer to be measurable, the large laser beam waists that would
take advantage of a larger number of molecules are often
unavailable in two-photon experiments. Second, miniaturization of the
decelerator helps in reducing the magnitude of stray fields. Finally,
miniaturization can be convenient in the case of a
Ramsey-configuration experiment, where mechanical stability to the
interferometric level is needed.

\begin{acknowledgments}
I gratefully acknowledge the fruitful discussion with Samuel A. Meek. 
\end{acknowledgments}

\bibliographystyle{gams-notit-nonumb}
\bibliography{coldmol}

\providecommand{\bysame}{\leavevmode\hbox to3em{\hrulefill}\thinspace}
\begin{thebibliography}{10}

\bibitem{Daw_Nature442p367_2006}
R.~Daw and J.~Finkelstein, \emph{Nature} \textbf{442} (2006), 367.

\bibitem{Morris_Astrobiology12p830_2012}
H.~C. Morris, M.~Damon, J.~Maule, L.~A. Monaco, and N.~Wainwright,
  \emph{Astrobiology} \textbf{12} (2012), 830.

\bibitem{Frisk_LabChip6p1504_2006}
T.~Frisk, D.~Ronnholm, W.~van~der Wijngaart, and G.~Stemme, \emph{Lab Chip}
  \textbf{6} (2006), 1504.

\bibitem{Heine_IntRevPhysChem34p1_2015}
N.~Heine and K.~R. Asmis, \emph{Int. Rev. Phys. Chem.} \textbf{34} (2015), 1.

\bibitem{Asmis_AccChemRes45p43_2012}
K.~R. Asmis and D.~M. Neumark, \emph{Acc. Chem. Res.} \textbf{45} (2012), 43.

\bibitem{Johnson_JChemPhys139p224305_2013}
C.~J. Johnson, J.~A. Fournier, C.~T. Wolke, and M.~A. Johnson, \emph{J. Chem.
  Phys.} \textbf{139} (2013), 224305.

\bibitem{Fournier_Science344p1009_2014}
J.~A. Fournier, C.~J. Johnson, C.~T. Wolke, G.~H. Weddle, A.~B. Wolk, and M.~A.
  Johnson, \emph{Science} \textbf{344} (2014), 1009.

\bibitem{Jiang_JPhysChemA115p11187_2011}
L.~Jiang, T.~Wende, P.~Claes, S.~Bhattacharyya, M.~Sierka, G.~Meijer,
  P.~Lievens, J.~Sauer, and K.~R. Asmis, \emph{J. Phys. Chem. A} \textbf{115}
  (2011), 11187.

\bibitem{Janssens_PhysRevLett96p233401_2006}
E.~Janssens, G.~Santambrogio, M.~Br\"{u}mmer, L.~W\"{o}ste, P.~Lievens,
  J.~Sauer, G.~Meijer, and K.~R. Asmis, \emph{Phys. Rev. Lett.} \textbf{96}
  (2006), 233401.

\bibitem{Metcalf_LaserCooling1999}
H.~J. Metcalf and P.~van~der Straten, \emph{Laser cooling and trapping},
  Springer, Berlin, 1999.

\bibitem{Fortagh_RevModPhys79p235_2007}
J.~Fort\'{a}gh and C.~Zimmermann, \emph{Rev. Mod. Phys.} \textbf{79} (2007),
  235.

\bibitem{Hansel_Nature413p498_2001}
W.~H\"{a}nsel, P.~Hommelhoff, T.~W. H\"{a}nsch, and J.~Reichel, \emph{Nature}
  \textbf{413} (2001), 498.

\bibitem{Schumm_NatPhysics1p57_2005}
T.~Schumm, S.~Hofferberth, L.~M. Andersson, S.~Wildermuth, S.~Groth,
  I.~Bar-Joseph, J.~Schmiedmayer, and P.~Kr\"uger, \emph{Nat. Physics}
  \textbf{1} (2005), 57.

\bibitem{Zoest_Science328p1540_2010}
T.~van Zoest, N.~Gaaloul, Y.~Singh, H.~Ahlers, W.~Herr, S.~T. Seidel,
  W.~Ertmer, E.~Rasel~M. Eckart, E.~Kajari, S.~Arnold, G.~Nandi, W.~P.
  Schleich, R.~Walser, A.~Vogel, K.~Sengstock, K.~Bongs, W.~Lewoczko-Adamczyk,
  M.~Schiemangk, T.~Schuldt, A.~Peters, T.~Ko\"onemann, H.~M\"untinga,
  C.~L\"ammerzahl, H.~Dittus, T.~Steinmetz, T.~W. H\"ansch, and J.~Reichel,
  \emph{Science} \textbf{328} (2010), 1540.

\bibitem{Ospelkaus_Nature476p181_2011}
C.~Ospelkaus, U.~Warring, Y.~Colombe, K.~R. Brown, J.~M. Amini, D.~Leibfried,
  and D.~J. Wineland, \emph{Nature} \textbf{476} (2011), 181.

\bibitem{Gring_Science337p1318_2012}
M.~Gring, M.~Kuhnert, T.~Langen, T.~Kitagawa, B.~Rauer, M.~Schreitl, I.~Mazets,
  D.~Adu Smith, E.~Demler, and J.~Schmiedmayer, \emph{Science} \textbf{337}
  (2012), 1318.

\bibitem{Baron_Science343p269_2014}
J~Baron, W.~C Campbell, D~Demille, J.~M Doyle, G~Gabrielse, Y.~V Gurevich, P.~W
  Hess, N.~R Hutzler, E~Kirilov, I~Kozyryev, B.~R O'leary, C.~D Panda, M.~F
  Parsons, E.~S Petrik, B~Spaun, A.~C Vutha, and A.~D West, \emph{Science}
  \textbf{343} (2014), 269.

\bibitem{Daussy_PhysRevLett83p1554_1999}
C.~Daussy, T.~Marrel, A.~Amy-Klein, C.~T. Nguyen, C.~J. Bord\'{e}, and
  C.~Chardonnet, \emph{Phys. Rev. Lett.} \textbf{83} (1999), 1554.

\bibitem{Salumbides_PhysRevD87p112008_2013}
E.~J. Salumbides, J.~C.~J. Koelemeij, J.~Komasa, K.~Pachucki, K.~S.~E. Eikema,
  and W.~Ubachs, \emph{Phys. Rev. D} \textbf{87} (2013), 112008.

\bibitem{Salumbides_PhysRevLett107p043005_2011}
E.~J. Salumbides, G.~D. Dickenson, T.~I. Ivanov, and W.~Ubachs, \emph{Phys.
  Rev. Lett.} \textbf{107} (2011), 043005.

\bibitem{Daussy_PhysRevLett98p250801_2007}
C.~Daussy, M.~Guinet, A.~Amy-Klein, K.~Djerroud, Y.~Hermier, S.~Briaudeau,
  C.~J. Bord\'{e}, and C.~Chardonnet, \emph{Phys. Rev. Lett.} \textbf{98}
  (2007), 250801.

\bibitem{Shelkovnikov_PRL100p150801_2008}
A.~Shelkovnikov, R.~Butcher, C.~Chardonnet, and A.~Amy-Klein, \emph{Phys. Rev.
  Lett.} \textbf{100} (2008), 150801.

\bibitem{Truppe_NatureComm4p2600_2013}
S.~Truppe, R.J. Hendricks, S.K. Tokunaga, H.J. Lewandowski, M.G. Kozlov,
  C.~Henkel, E.A. Hinds, and M.R. Tarbutt, \emph{Nature Comm.} \textbf{4}
  (2013), 2600.

\bibitem{Bagdonaite_Science339p46_2013}
J.~Bagdonaite, P.~Jansen, C.~Henkel, H.~L. Bethlem, K.~M. Menten, and
  W.~Ubachs, \emph{Science} \textbf{339} (2013), 46.

\bibitem{Salumbides_PhysRevLett101p223001_2008}
E.~J. Salumbides, D.~Bailly, A.~Khramov, A.~L. Wolf, K.~S.~E. Eikema,
  M.~Vervloet, and W.~Ubachs, \emph{Phys. Rev. Lett.} \textbf{101} (2008),
  223001.

\bibitem{DeMille_PhysRevLett88p067901_2002}
D.~DeMille, \emph{Phys. Rev. Lett.} \textbf{88} (2002), 067901.

\bibitem{Andre_NatPhys2p636_2006}
A.~Andr\'e, D.~DeMille, J.~M. Doyle, M.~D. Lukin, S.~E. Maxwell, P.~Rabl, R.~J.
  Schoelkopf, and P.~Zoller, \emph{Nat. Phys} \textbf{2} (2006), 636.

\bibitem{Goral_PhysRevLett88p170406_2002}
K.~G\'{o}ral, L.~Santos, and M.~Lewenstein, \emph{Phys. Rev. Lett.} \textbf{88}
  (2002), 170406.

\bibitem{Micheli_NaturePhys2p341_2006}
A.~Micheli, G.~K. Brennen, and P.~Zoller, \emph{Nature Phys.} \textbf{2}
  (2006), 341.

\bibitem{Krems_PhysChemChemPhys10p4079_2008}
R.~V. Krems, \emph{Phys. Chem. Chem. Phys.} \textbf{10} (2008), 4079.

\bibitem{deMiranda_NaturePhys7p502_2011}
M.~H.~G. de~Miranda, A.~Chotia, B.~Neyenhuis, D.~Wang, G.~Qu\'{e}m\'{e}ner,
  S.~Ospelkaus, J.~L. Bohn, J.~Ye, and D.~S. Jin, \emph{Nature Phys.}
  \textbf{7} (2011), 502.

\bibitem{Avdeenkov_PhysRevA66p052718_2002}
A.~V. Avdeenkov and J.~L. Bohn, \emph{Phys. Rev. A} \textbf{66} (2002), 052718.

\bibitem{Tscherbul_JChemPhys129p034112_2008}
T.~V. Tscherbul and R.~V. Krems, \emph{J. Chem. Phys.} \textbf{129} (2008),
  034112.

\bibitem{Shuman_Nature467p820_2010}
E.~S. Shuman, J.~F. Barry, and D.~DeMille, \emph{Nature} \textbf{467} (2010),
  820.

\bibitem{Barry_PhysRevLett108p103002_2012}
J.~F. Barry, E.~S. Shuman, E.~B. Norrgard, and D.~DeMille, \emph{Phys. Rev.
  Lett.} \textbf{108} (2012), 103002.

\bibitem{Zhelyazkova_PhysRevA89p053416_2014}
V.~Zhelyazkova, A.~Cournol, T.~E. Wall, A.~Matsushima, J.~J. Hudson, E.~A.
  Hinds, M.~R. Tarbutt, and B.~E. Sauer, \emph{Phys. Rev. A} \textbf{89}
  (2014), 053416.

\bibitem{Weinstein_Nature395p148_1998}
J.~D. Weinstein, R.~deCarvalho, T.~Guillet, B.~Friedrich, and J.~M. Doyle,
  \emph{Nature} \textbf{395} (1998), 148.

\bibitem{Meek_NewJPhys11p055024_2009}
S.~A. Meek, H.~Conrad, and G.~Meijer, \emph{New J. Phys.} \textbf{11} (2009),
  055024.

\bibitem{Meek_PhysRevLett100p153003_2008}
S.~A. Meek, H.~L. Bethlem, H.~Conrad, and G.~Meijer, \emph{Phys. Rev. Lett.}
  \textbf{100} (2008), 153003.

\bibitem{Bethlem_Nature406p491_2000}
H.~L. Bethlem, G.~Berden, F.~M.~H. Crompvoets, R.~T. Jongma, A.~J.~A. van Roij,
  and G.~Meijer, \emph{Nature} \textbf{406} (2000), 491.

\bibitem{Rieger_PRL95p173002_2005}
T.~Rieger, T.~Junglen, S.~Rangwala, P.~W.~H. Pinkse, and G.~Rempe, \emph{Phys.
  Rev. Lett.} \textbf{95} (2005), 173002.

\bibitem{Bethlem_IntRevPhysChem22p73_2003}
H.~L. Bethlem and G.~Meijer, \emph{Int. Rev. Phys. Chem.} \textbf{22} (2003),
  73.

\bibitem{Meerakker_ChemRev112p4828_2012}
S.~Y.~T. van~de Meerakker, HL~Bethlem, N~Vanhaecke, and G~Meijer, \emph{Chem.
  Rev.} \textbf{112} (2012), 4828.

\bibitem{Opat_ApplPhysB54p396_1992}
G.~I. Opat, S.~J. Wark, and A.~Cimmino, \emph{Appl. Phys. B} \textbf{54}
  (1992), 396.

\bibitem{Schulze_PhysRevLett93p020406_2004}
S.~A. Schulze, H.~L. Bethlem, J.~van Veldhoven, J.~K\"upper, H.~Conrad, and
  G.~Meijer, \emph{Phys. Rev. Lett.} \textbf{93} (2004), 020406.

\bibitem{GonzalezFlorez_PhysChemChemPhys13p18830_2011}
A.~I.~Gonz\'alez Fl\'orez, S.~A. Meek, H.~Haak, H.~Conrad, G.~Santambrogio, and
  G.~Meijer, \emph{Phys. Chem. Chem. Phys.} \textbf{13} (2011), 18830.

\bibitem{Englert_PhysRevLett107p263003_2011}
B.~G.~U. Englert, M.~Mielenz, C.~Sommer, J.~Bayerl, M.~Motsch, P.~W.~H. Pinkse,
  G.~Rempe, and M.~Zeppenfeld, \emph{Phys. Rev. Lett.} \textbf{107} (2011),
  263003.

\bibitem{Marx_PhysRevLett111p243007_2013}
S.~Marx, D.~Adu~Smith, M.~J. Abel, T.~Zehentbauer, G.~Meijer, and
  G.~Santambrogio, \emph{Phys. Rev. Lett.} \textbf{111} (2013), 243007.

\bibitem{Petrich_PhysRevLett74p3352_1995}
W.~Petrich, M.~H. Anderson, J.~R. Ensher, and E.~A. Cornell, \emph{Phys. Rev.
  Lett.} \textbf{74} (1995), 3352.

\bibitem{Kirste_PhysRevA79p051401R_2009}
M.~Kirste, B.~G. Sartakov, M.~Schnell, and G.~Meijer, \emph{Phys. Rev. A}
  \textbf{79} (2009), 051401(R).

\bibitem{Meek_PhysRevA71p065402_2005}
S.~A. Meek, E.~R.~I. Abraham, and N.~E. Shafer-Ray, \emph{Phys. Rev. A}
  \textbf{71} (2005), 065402.

\bibitem{Meek_Science324p1699_2009}
S.~A. Meek, H.~Conrad, and G.~Meijer, \emph{Science} \textbf{324} (2009), 1699.

\bibitem{Meek_PhysRevA83p033413_2011}
S.~A. Meek, G.~Santambrogio, B.~Sartakov, H.~Conrad, and G.~Meijer, \emph{Phys.
  Rev. A} \textbf{83} (2011), 033413.

\bibitem{Abel_MolecularPhysics110p1829_2012}
M.~J. Abel, S.~Marx, G.~Meijer, and G.~Santambrogio, \emph{Molecular Physics}
  \textbf{110} (2012), 1829.

\bibitem{Santambrogio_ChemPhysChem12p1799_2011}
G.~Santambrogio, S.~A Meek, M.~J. Abel, L.~M. Duffy, and G.~Meijer,
  \emph{ChemPhysChem} \textbf{12} (2011), 1799.

\bibitem{Ashfold_AnnuRevPhysChem45p57_1994}
M.~N.~R. Ashfold and J.~D. Howe, \emph{Annu. Rev. Phys. Chem.} \textbf{45}
  (1994), 57.

\bibitem{Bulleid_PhysRevA86p021404R_2012}
N.~E. Bulleid, R.~J. Hendricks, E.~A. Hinds, S.~A. Meek, G.~Meijer,
  A.~Osterwalder, and M.~R. Tarbutt, \emph{Phys. Rev. A} \textbf{86} (2012),
  021404R.

\end{thebibliography}

\end{document}